# Fabrication and characterization of iron pnictide wires and bulk materials through the powder-in-tube method

Yanwei Ma*, Zhaoshun Gao, Yanpeng Qi, Xianping Zhang, Lei Wang, Zhiyu Zhang, Dongliang Wang

Key Laboratory of Applied Superconductivity, Institute of Electrical Engineering,

Chinese Academy of Sciences, P. O. Box 2703, Beijing 100190, China

**Abstract:** The recent discovery of superconductivity in the iron based superconductors with very high upper critical fields presents a new possibility for practical applications, but fabricating fine-wire is a challenge because of mechanically hard and brittle powders and the toxicity and volatility of arsenic. In this paper, we report the synthesis and the physical characterization of iron pnictide wires and bulks prepared by the powder-in-tube method (PIT). A new class of high-$T_c$ iron pnictide composite wires, such as $LaFeAsO_{1-x}F_x$, $SmFeAsO_{1-x}F_x$ and $Sr_{1-x}K_xFeAs$, has been fabricated by the in situ PIT technique using Fe, Ta and Nb tubes. Microscopy and x-ray analysis show that the superconducting core is continuous, and retains phase composition after wire drawing and heat treatment. Furthermore, the wires exhibit a very weak $J_c$-field dependence behavior even at high temperatures. The upper critical field $H_{c2}(0)$ value can exceed 100 T, surpassing those of $MgB_2$ and all the low temperature superconductors and indicating a strong potential for applications requiring very high field. These results demonstrate the feasibility of producing superconducting pnictide composite wire. We also applied the one step PIT method to synthesize the iron-based bulks, due to its convenience and safety. In fact, by using this technique, we have successfully discovered superconductivity at 35 K and 15 K in $Eu_{0.7}Na_{0.3}Fe_2As_2$ and SmCoFeAsO compounds, respectively. These clearly suggest that the one-step PIT technique is unique and versatile and hence can be tailored easily for other rare earth derivatives of novel iron-based superconductors.

---

\* Corresponding author. Fax: +86-10-82547137.    E-mail: ywma@mail.iee.ac.cn (Y. Ma)

# 1. Introduction

The recently discovered LaFeAsO$_{1-x}$F$_x$ with a transition temperature (T$_c$) of 26 K represents a new class of high-T$_c$ superconductors (FeAs-1111 phase) [1]. Soon after the discovery, it was reported that the replacement of La by other rare-earth element (*RE*), such as Pr, Sm, Eu, Nd and Gd, significantly increases T$_c$ [2-6]. To date, the highest T$_c$ exceeds 50 K. Similar to the CuO-layered high-T$_c$ cuprates, the new compounds are also layered materials, with the superconductivity occurring in FeAs sheets, with the *RE*-O layers acting as a charge reservoir. Moreover, a simpler class of materials based on the BaFe$_2$As$_2$ parent compound (FeAs-122 structure) that does not have the *RE*-O spacer layers also exhibits superconductivity with a comparable T$_c$ to the 1111 materials [7].

Rather high upper critical fields over 100 T have been reported for iron oxypnictide superconductors [8-10]. In particular, a recent study found a very large $H_{c2}^{ab}$ (0) ≈ 304 T in a NdFeAsO$_{0.82}$F$_{0.18}$ single crystal [11]. Such critical field properties can be competitive with those of A15, MgB$_2$ and even high-T$_c$ cuprate conductors, suggesting that iron pnictides have great potential for high-field applications above 30 K, where conventional superconductors cannot play a role owing to their low T$_c$s.

Many groups have now fabricated oxypnictide polycrystalline bulks by high-pressure synthesis technique and conventional solid state reaction methods (two or one step method) [2-6, 10, 12], and have reported structural and magnetically derived $J_c$ properties. Intergrain critical currents of the order of 10$^3$-10$^4$ A/cm$^2$ have been reported in the temperature range between 5 and 30 K in bulk samples [9, 10, 12-16]. Recently, even larger critical current densities with weak field dependence have been reported in the 122 phase Ba$_{1-x}$K$_x$Fe$_2$As$_2$ system [17]. It is expected that further optimization of the synthesis process will raise these $J_c$ values closer to the intrinsic intragrain current density of the oxypnictide phase, which has been evaluated to be of the order of 10$^6$ A/cm$^2$ [9, 10].

The properties of iron-based superconductors are increasing rather quickly, suggesting that this system seems much easier to work with than "high T$_c$" cuprates and $MgB_2$. From the viewpoint of practical applications, such as magnets and cables, the next step is to make the material in wire and tape forms. However, it is really a challenge to fabricate pnictide wires, because these materials are mechanically hard and brittle and therefore not easy to draw into the desired wire geometry. On the other hand, the use of proper metal cladding is another critical issue because of the strong chemical reactivity of the pnictide at the high annealing temperatures. Recently, our group produced the first $LaFeAsO_{0.9}F_{0.1}$ and $SmFeAsO_{1-x}F_x$ wires by the powder-in-tube (PIT) method and they exhibited $J_c$ values of $\sim 4\times 10^3$ A/cm$^2$ at 5 K [18, 19]. It should be noted that this process is also suitable for preparing iron-based bulk superconductors after a small modification [10].

In the light of these early results and of the rapidly growing literature dealing with the preparation of iron pnictide compounds, the feasibility of fabricating powder-in-tube processed pnictide wires needs to be explored in detail. The aim of this paper is to present the preparation procedures and the most recent results and developments of iron pnictide wires and bulks made by the PIT technique.

**2. Methods for the synthesis of iron pnictide bulk superconductors**

Up to now, the new family of iron based superconductors is only prepared by a limited number of research groups, due to the very complicated synthesis route and the toxicity and volatility of arsenic. Actually, so far most reaearch teams in the world analyzed this new high-temperature superconductor by employing samples obtained from China.

**2.1 Two-step sealed vacuum quartz-tube synthesis**

A conventional way of producing pnictide polycrystalline samples is the so-called two-step sealed vacuum quartz-tube synthesis process, which is currently used in a large number of laboratories. This new superconductor contains arsenic, a toxic element, and its chemical characteristics are hard to manipulate under laboratory conditions, because evacuated quartz tubes have to be employed. Sometime explosive accidents occur due to over pressurization. Another disadvantage is that fluorine readily reacts with quartz. The typical preparation process is as follows [1, 2]:

Starting materials are high purity $RE$As ($RE$: rare earth), $RE$F$_3$, Fe and $Fe_2O_3$.

The binary compounds $RE$As and FeAs were pre-sintered in evacuated quartz tubes using Fe powder, $RE$ and As pieces at 500°C for 5-10 hours and 900°C for 10-24 hours. The raw materials were thoroughly grounded in Ar and pressed into pellets. The pellets were wrapped in Ta foil, sealed in an evacuated quartz tube, and then, sintered at 1100-1200°C for 40 hours.

**2.2 High-pressure high temperature synthesis**

The superconducting iron pnictide SmFeAsO$_{1-x}$F$_x$ bulks with the highest Tc= 55 K were prepared by a high pressure (HP) synthesis method [4, 6]. SmAs powder (pre-sintered) and As, Fe, Fe$_2$O$_3$, FeF$_2$ powders with purity better than 99.99% were mixed together according to the nominal ratio, then ground thoroughly and pressed into small pellets. The pellets were sealed in boron nitride crucibles and sintered in a high pressure synthesis apparatus under a high pressure of 6 GPa and high temperature of 1250°C for 2 hours.

So far, the HP method seems very efficient for synthesizing pnictide compounds, due to the F and As being enclosed within the pressure capsule, but because of the short sintering time it is difficult to avoid the existence of some unreacted components, such as FeAs and others. Another noted thing is that the HP apparatus is usually quite expensive and also is difficult to maintain.

**2.3 One-step PIT process**

In order to simplify the fabrication process and to avoid the accidental explosions, we have proposed one simple method for fabrication of the new iron based superconductors [10]. In this process, we modified the well-known powder-in-tube (PIT) technique commonly applied for fabrication of wires and tapes, and wire-shaped bulk samples can be obtained. Compared to the above mentioned two processes, this PIT method involves a simple procedure: thoroughly mixing all the starting materials one time, packing the powders in a metal tube, and sintering. The major merits of the fabrication method are safety and convenience.

By using this PIT method, we have successfully synthesized SmFeAsO$_{0.8}$F$_{0.2}$ bulk samples. Starting materials, such as high-purity Sm, As pieces, SmF$_3$, Fe and Fe$_2$O$_3$ powders, were thoroughly mixed and loaded into a Ta or Nb tube under an Ar atmosphere, and both tube ends were tightly closed by Nb plugs. The tubes were then swaged, sealed in an Fe tube by welding, and annealed under Ar at 1160°C for 40 hours. The sintered samples were taken out by breaking the Fe and sheath tubes.

The resulting SmFeAsO$_{0.8}$F$_{0.2}$ bulks show promising results; superconducting properties of the samples are quite comparable to those of samples made by the two-step sealed vacuum quartz-tube fabrication route. We can observe a sharp transition with an onset temperature $T_c$ = 51 K and zero resistance at 47 K, indicating the good quality of our samples, as shown in Figure 1. It should be noted that $T_c$ increased with increasing fluorine content, and reached its maximum of 54.6 K at $x$=0.3 [10].

**3. Characterization methodology**

Phase identification and crystal structure investigation were carried out using x-ray diffraction (XRD) with a Philips X'Pert PRO system (Cu Kα radiation, λ=1.5406Å). Microstructural observations were performed on a scanning electron microscopy (SEM) equipped with energy dispersive spectroscopy (EDS). Resistivity measurements were carried out by the conventional four-point-probe method using a Quantum Design PPMS. Magnetization of the samples was measured by a SQUID magnetometer (Quantum Design, MPMS XL-7). The 10% and 90% points on the resistive transition curves were used to define the H$_{irr}$ and H$_{c2}$, respectively. The critical current density was calculated by using the Bean model.

**4. Discovery of new pnictide compounds through one-step PIT process**

It seems that the one-step PIT method suggested by our group is quite effective and fast for exploring new iron-based superconductors, since we never require the process of pre-sintering *RE*As and FeAs as the first step. In fact, recently, we have successfully discovered superconductivity at 34.7 K and 15.2 K in new Eu$_{0.7}$Na$_{0.3}$Fe$_2$As$_2$ and Co-doped SmFeAsO compounds, respectively [20, 23], by using the one-step PIT technique.

We mixed staring materials of Sm, Co$_2$O$_3$, Fe, Fe$_2$O$_3$ and As together to synthesize a Co-doped SmFeAsO compound, and found superconductivity can be induced by partial substitution of Fe by transition metal element Co in the superconducting-active $Fe_2As_2$ layers [20], which is in contrast to high-$T_c$ cuprates. As we know, the parent compound SmFeAsO itself is not superconducting but shows an antiferromagnetic order near 150 K. With Co doping in the FeAs planes, antiferromagnetic order is destroyed and superconductivity occurs at 15.2 K (see Fig.1). Similar to Co-doped LaFeAsO and CaFeAsF [21, 22], the $SmFe_{1-x}Co_xAsO$ system appears to tolerate considerable disorder in the FeAs planes. This result is important, suggesting a different mechanism for cuprate superconductors compared to the iron-based arsenide ones. It should be noted that this appears to be the first discovery of Co-doped SmFeAsO superconductor.

On the other hand, $EuFe_2As_2$ is a member of the ternary iron arsenide family. Like $BaFe_2As_2$ and $SrFe_2As_2$, $EuFe_2As_2$ is a poor metal, with a spin density wave anomaly at about 200 K. Based on the one-step PIT method, we observed superconductivity in $Eu_{0.7}Na_{0.3}Fe_2As_2$ by partial substitution of the europium site with sodium [23]. $ThCr_2Si_2$ tetragonal structure, as expected for $EuFe_2As_2$, is formed as the main phase for the composition $Eu_{0.7}Na_{0.3}Fe_2As_2$. Resistivity measurement reveals that the transition temperature $T_c$ of this compound is as high as 34.7 K, as shown in Fig.1, which is comparable to the $T_c$ of $Eu_{0.5}K_{0.5}Fe_2As_2$ [24]. The rate of $T_c$ suppression with applied magnetic field is ~3.9 T / K, giving an extrapolated zero-temperature upper critical field of 90 T, which is suitable for high field applications. This is very similar to the properties of the $Ba_{1-x}K_xFe_2As_2$.

Our results clearly demonstrated that the one-step PIT synthesis technique is unique and versatile and hence can be tailored easily for other rare earth derivatives of pnictide superconductors.

**5. Preparation of iron pnictide wires by the PIT method**

Iron pnictides are a relatively tough and hard phases, and thus cannot be plastically deformed. The only way to obtain a filamentary configuration is to start with powders that are packed in metallic tubes, in analogy to high $T_c$ Bi-cuprate superconductor. Therefore, we choose the powder-in-tube (PIT) method to fabricate the iron based superconducting wires. The PIT process takes advantage of the low material costs and the deformation techniques employed are relatively simple.

The choice of the metallic sheath has been reduced to those elements or alloys showing little or no reaction with pnictides at high temperatures. So far, Nb, Ta and Fe have been found to be the appropriate sheath material for wire fabrication, showing little reaction with pnictide during the final heat treatment.

Iron pnictide wires are fabricated by filling stoichiometric amounts of *RE* (*RE*: rare earth), As, *RE*F$_3$, Fe and Fe$_2$O$_3$ powder particles into metal tubes under Ar atmosphere. The tubes are then swaged and drawn to wires of ~2.2 mm in diameter and finally given a heat treatment under Ar. The details of the fabrication process are described elsewhere [18-19]. The wire processing steps are shown schematically in Figure 2.

Figure 3 shows a photograph of the final Ta-sheathed SmFeAsO$_{1-x}$F$_x$ pnictide wires. The transverse and longitudinal cross-sections of a sintered SmFeAsO$_{1-x}$F$_x$ wire are quite uniform and are shown in Fig. 3a and 3b. A reaction layer with a thickness 10~30 μm between the superconductor core and the Ta tube was clearly observed after 45 h at 1180°C [19]. EDS/SEM analysis detected a small amount of Ta in the superconductor core near the Ta sheath. Since the heating temperatures for 122 phase pnictide are quite low, this reaction layer is expected to be very small.

**6. Properties of iron pnictide wires**

Figure 4 presents the typical XRD pattern of SmFeAsO$_{0.65}$F$_{0.35}$ wires after peeling off the Ta sheath materials. The sample consists of a main superconducting phase, but some impurity phases are also detected. Such impurity phases can be reduced by optimizing the heating process and stoichiometry ratio of the starting materials. The lattice parameters are a = 3.9310 Å, c = 8.5221 Å. The a-axis lattice constant decreases steadily with increasing substitution of F$^-$ for O$^{2-}$ [19]. These data are in good agreement with results for bulk samples synthesized by the two step method [2].

SEM images of several kinds of pnictide superconducting wires are shown in Figure 5. Clearly, well-developed grains can be seen in all samples, however, some voids and poor intergrain connections are present (see Fig.5a and 5b). In other words, weak links exist at the grain boundaries of this sample, and thus introduce a strong limitation to the flow of current. We should note that the grains have plate-like morphology, especially in Nb-sheathed SmFeAsO wire samples (Fig.5d), which strongly supports the layered nature in this type of superconductor, very similar to what has been observed in Bi-based cuprates. The average lateral grain size was found to be around 10 um. The EDX results suggested that superconducting grains are compositionally homogeneous, at least within the limits of SEM-EDX analysis. At present the main difficulty when forming the pnictide phase inside the wires is to obtain samples of sufficient homogeneity and high density, since the packing process has to be done in a glove box filled with inert gas. Thus the density of the samples is generally too low.

Figure 6 shows typical temperature dependence of resistance for the LaFeAsO$_{0.9}$F$_{0.1}$, Sr$_{0.5}$K$_{0.5}$FeAs and SmFeAsO$_{0.65}$F$_{0.35}$ wires using various sheaths. All the samples show metallic (d$\rho$/dT > 0) behavior in all the temperature range. From Fig. 6, sharp superconducting transitions are observed. The onset superconducting transition temperatures, T$_c$(onset), for these pnictide wires are 25, 32 and 52 K, respectively, which are a bit lower than those of the corresponding bulks. DC susceptibility data also indicated very sharp superconducting transitions for the above three wires. The diamagnetic superconducting shielding fractions for all these wire samples are large enough to constitute bulk superconductivity.

The upper critical field, $H_{c2}$, is usually derived from magneto-resistivity measurements, taking the onset temperatures of the resistive transition. Figure 7 shows the temperature dependence of $H_{c2}$ and $H_{irr}$ with magnetic field for the SmFeAsO$_{0.65}$F$_{0.35}$ wire after peeling away the Ta sheath. It is clear that the curve for $H_{c2}$ (T) is very steep with a slope of − d$H_{c2}$/dT|$_{Tc}$ = 2.74 T / K. The calculated $H_{c2}$(0) values were over 120 T in the SmFeAsO$_{0.65}$F$_{0.35}$ wire [19]. It is also evident that the irreversibility field is higher than those of MgB$_2$ and all the low temperature superconductors. These high values of $H_{c2}$ and $H_{irr}$ indicate that this new superconductor has potential utility for application in very high fields.

Magnetization loops with a strong ferromagnetic background were observed in both SmFeAsO and LaFeAsO wire samples, which are believed to be contributed by the unreacted Fe or $Fe_2O_3$ impurity phases [9, 10]. The same behavior was also found in the oxypnictide bulks prepared by the HP method [15]. In contrast, we could not see the ferromagnetic background for SrKFeAs wires [25], indicating that high purity $AFe_2As_2$ superconductor may be easily synthesized due to very low heating temperature and no oxygen.

Figure 8 presents the $J_c(H)$ values at 5 K for the bar and powder forms of the $SmFeAsO_{1-x}F_x$ samples [19]. The global $J_c$s were calculated on the basis of $J_c = 20\Delta M/a(1-a/3b)$, where $\Delta M$ is the height of the magnetization loop and $a$ and $b$ are the dimensions of the sample perpendicular to the magnetic field, $a < b$. Intragrain $J_c$ was also evaluated on the basis of $J_c = 30\Delta M/<R>$ through magnetic measurements after grinding the superconducting core into powder. $<R>$ is the average grain size, which is about 10 μm measured by SEM. It can be seen that the $J_c$ based on the sample size is much lower than what should exist in individual grains. The $J_c$ value at 5 K for the $SmFeAsO_{0.65}F_{0.35}$ powder is $2\times10^5$ A/cm$^2$ with a very weak dependence on field, which shows that the $SmFeAsO_{0.65}F_{0.35}$ has a fairly large pinning force in the grain. The intragrain $J_c$ at 20-40 K also exhibits large current carrying ability and better $J_c$-field performance [26] compared to that of $MgB_2$ with $J_c$ dropping very quickly at 20 and 30 K and even at low fields. However, the global $J_c$ value of $\sim4\times10^3$ A/cm$^2$ at 5 K obtained for the SmFeAsO samples is significantly lower than that seen in random bulks of $MgB_2$ which generally attained $10^6$ A/cm$^2$ at 4.2 K.

From the very low magnetic $J_c$, we can see that the current path was greatly reduced by the porosities, multiphase composition and interfacial reaction between the sheath and the superconducting core. In order to get transport $J_c$ values, these issues must eventually be solved. It is believed that improvement in the $J_c$ properties requires optimization of processing parameters to achieve high grain alignment in analogy to high $T_c$ Bi-based cuprates.

## 7. Conclusions

In summary, high-$T_c$ iron based superconducting wires have been fabricated using a powder-in-tube method. We find high upper critical field values and a very weak $J_c$ field dependence in the wires obtained, presenting the possibility of using such wires for applications. The global $J_c$ values are currently small, but as of yet very little effort has been put into optimizing $J_c$. One of the critical problems encountered in the fabrication of the present composite wire is the low density of the superconducting core since the packing process was performed in a glove-box filled with inert gas, therefore, improving the packing density and reducing the impurity phases should be the next step. The present work explicitly proves the feasibility of processing pnictide superconductors into a composite wire form which is useful for potential applications. In addition, we also demonstrate that the one step PIT method suggested by our group is more convenient, more efficient and safer for synthesizing the iron-based bulk samples.

**Acknowledgements**

The authors thank Profs. Zizhao Gan, Haihu Wen, Liye Xiao and Liangzhen Lin for their help and useful discussion. This work was partly supported by the Natural Science Foundation of China (contract nos 50572104 and 50777062) and National '973' Program (grant no. 2006CB601004).

**Captions**

Figure 1 Temperature dependence of resistivity for bulk samples made by the one-step PIT method. (a) SmFeAsO$_{0.8}$F$_{0.2}$, (b) Eu$_{0.7}$Na$_{0.7}$Fe$_2$As$_2$Eu, (c) SmFe$_{0.85}$Co$_{0.15}$AsO. The inset shows the enlarged view near the transition temperature.

Figure 2 Schematic view of the deformation steps for the fabrication of an iron pnictide wire by the PIT technique.

Figure 3 Photograph of the SmFeAsO$_{1-x}$F$_x$ wires. SEM images for a typical transverse (a) and a longitudinal (b) cross-section of the wire after heat treatment [19].

Figure 4 XRD patterns of SmFeAsO$_{0.65}$F$_{0.35}$ wires after peeling away the Ta sheath. The impurity phases of SmAs and SmOF are marked by * and #, respectively.

Figure 5 Typical SEM micrographs of the fractured superconducting core layers of pnictide wires. (a) Fe/LaFeAsO$_{0.9}$F$_{0.1}$ wire, (b) Ta/SmFeAsO$_{0.7}$F$_{0.3}$ wire; High magnification images of SmFeAsO$_{0.7}$F$_{0.3}$ wires with (c) Ta sheath, and (d) Nb sheath.

Figure 6 Temperature dependence of resistance for different iron pnictide wires. (a) Fe/LaFeAsO$_{0.9}$F$_{0.1}$, (b) Nb/Sr$_{0.5}$K$_{0.5}$FeAs, (c) Ta/SmFeAsO$_{0.65}$F$_{0.35}$.

Figure 7 Temperature dependence of $H_{c2}$ and $H_{irr}$ for the SmFeAsO$_{0.65}$F$_{0.35}$ wires, determined from 90% and 10% points on the resistive transition curves.

Figure 8 Magnetic field dependence of $J_c$ at 5 K for the bar and powder of SmFeAsO$_{1-x}$F$_x$ wires [19].

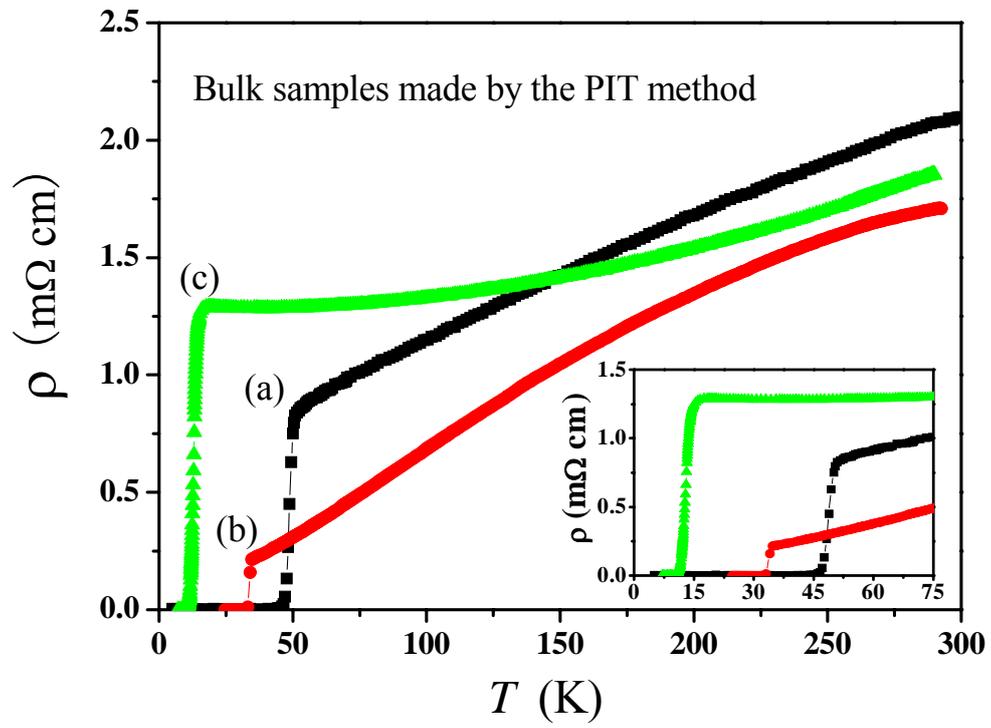

Figure 1   Ma et al.

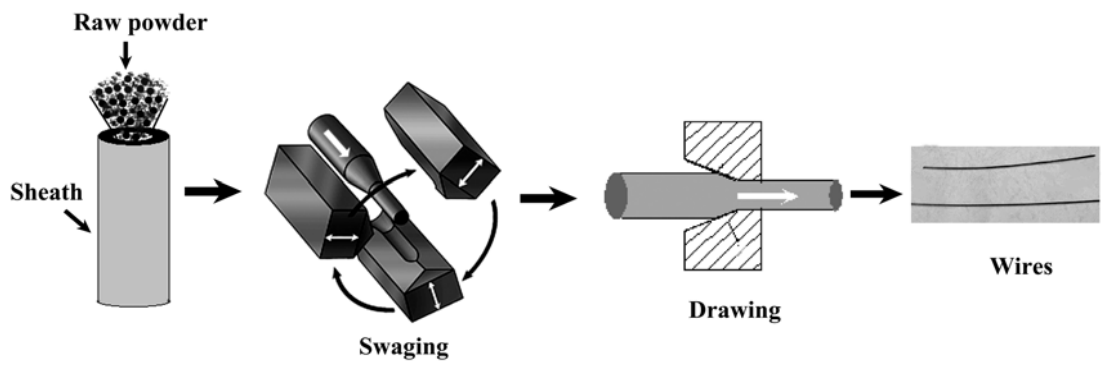

Figure 2    Ma et al.

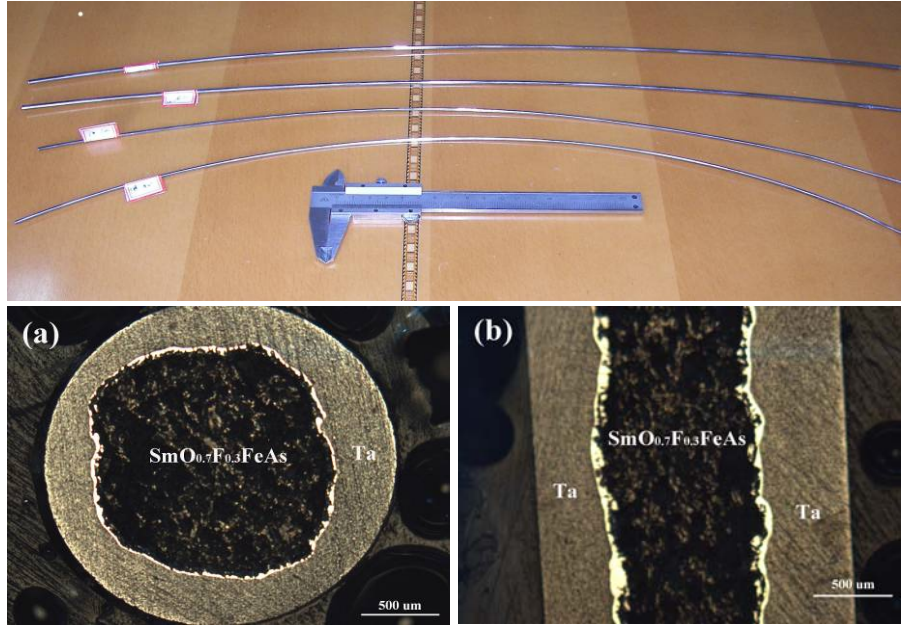

Figure 3    Ma et al.

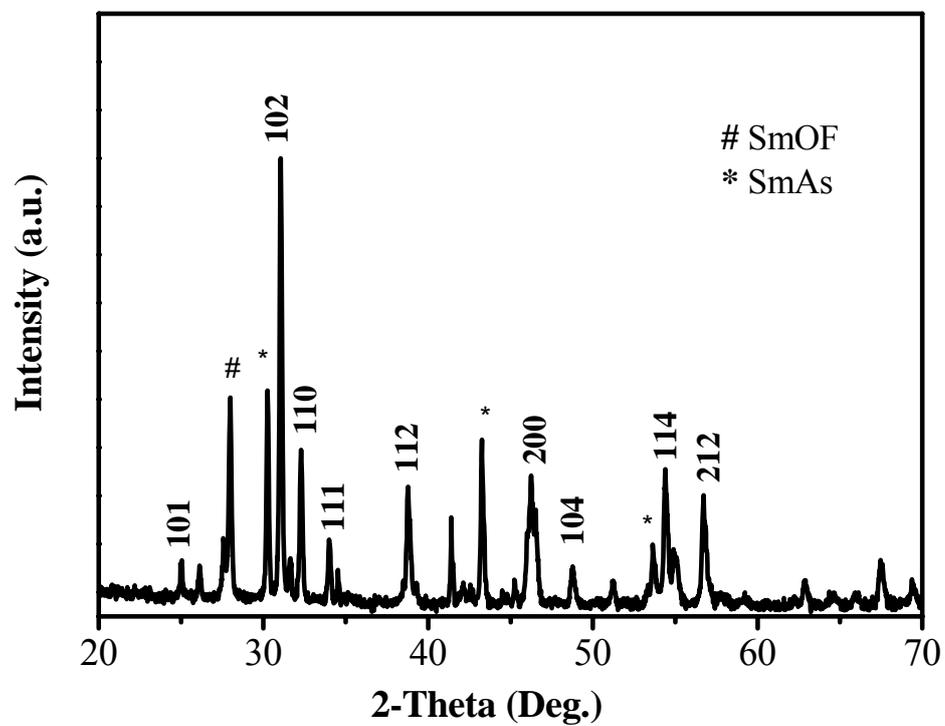

Figure 4    Ma et al.

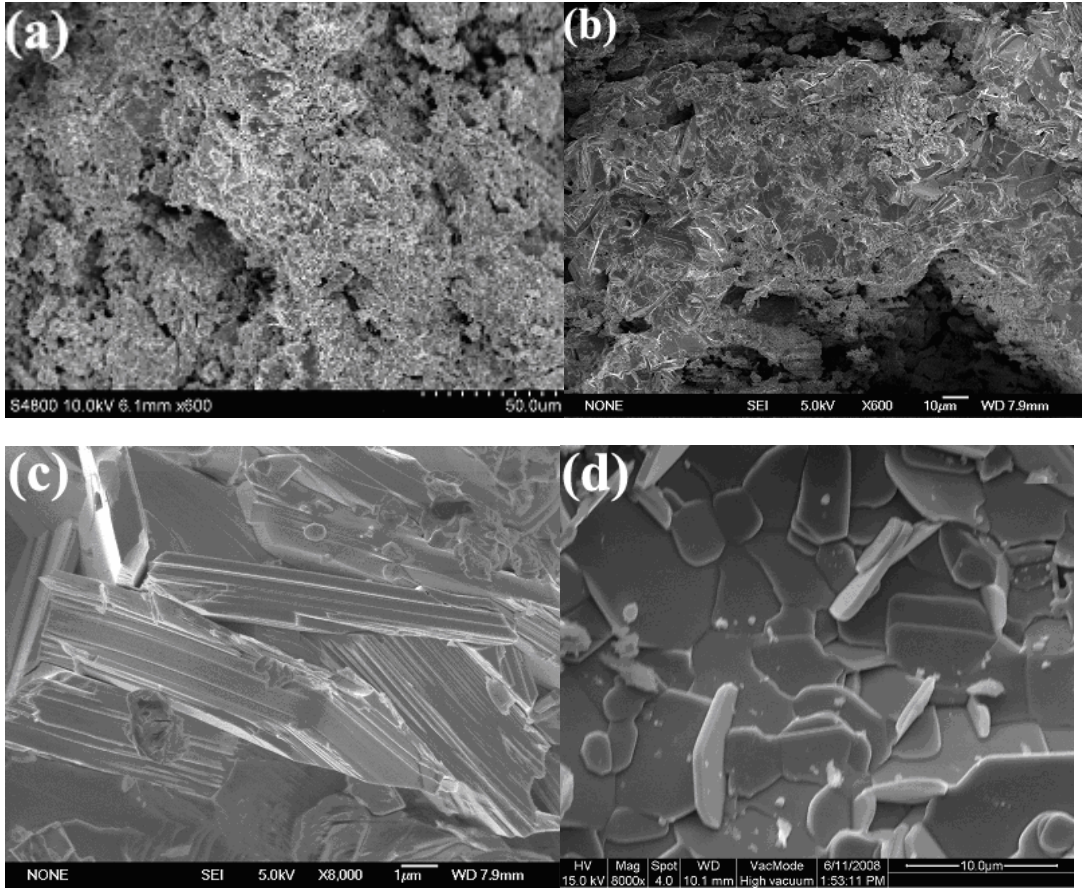

Figure 5  Ma et al.

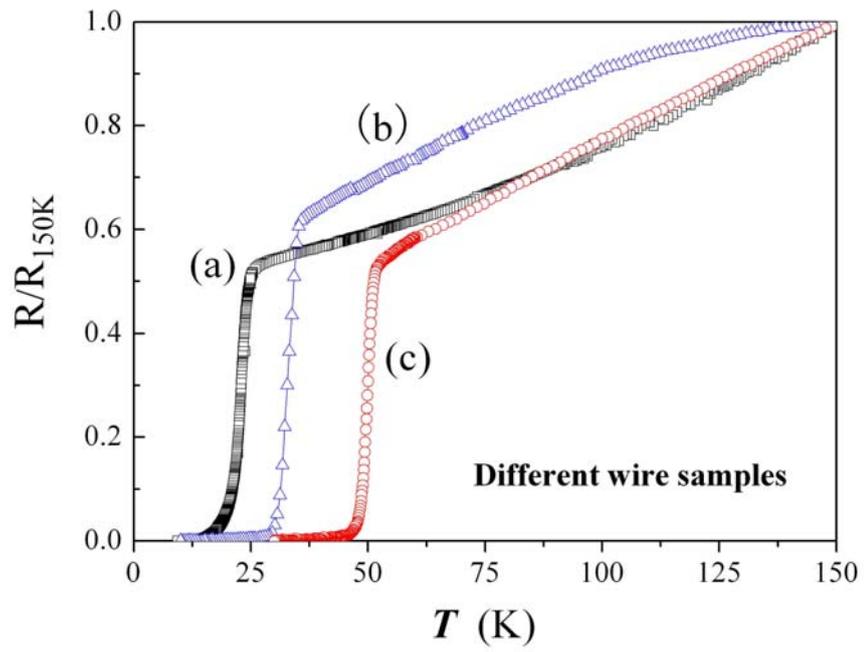

Figure 6　Ma et al.

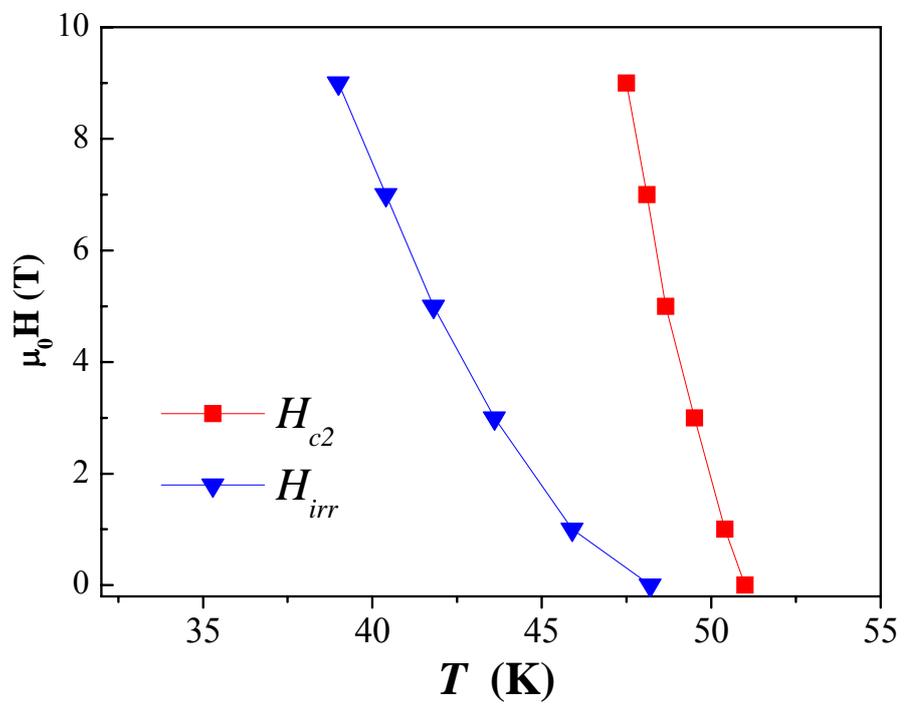

Figure 7    Ma et al.

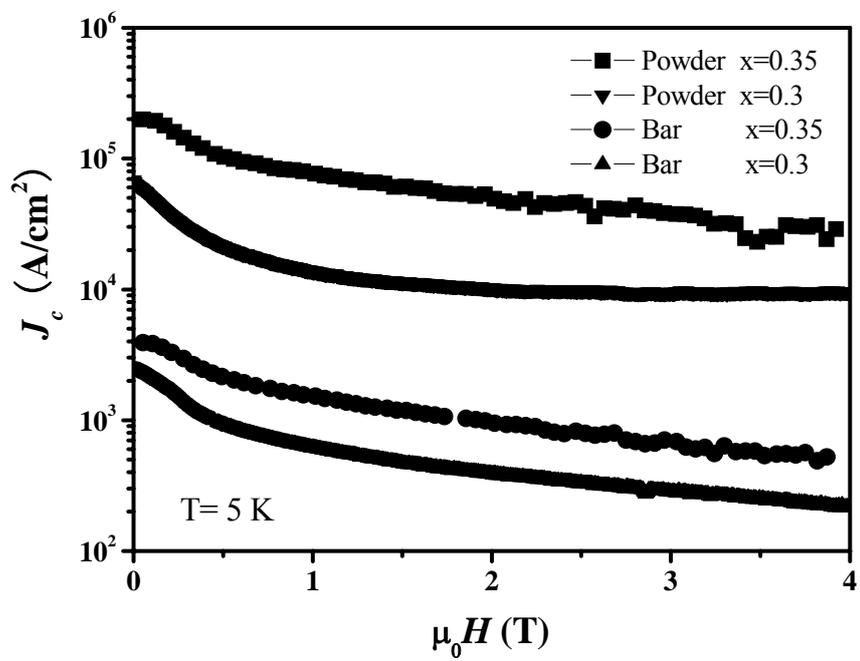

Figure 8    Ma et al.